**Universal temperature-dependent electrical resistivity in actinides**

Evgeny F. Talantsev[1,2*]

[1] M.N. Mikheev Institute of Metal Physics, Ural Branch, Russian Academy of Sciences, 18 S. Kovalevskaya St, Ekaterinburg 620108 Russia

[2] Nanotech Centre, Ural Federal University, 19 Mira St., Ekaterinburg 620002 Russia

*evgeny.talantsev@imp.uran.ru

## Abstract

Temperature-dependent electrical resistivity ρ($T$) is one of the most common types of experimental data analysed in condensed matter physics. For one group of pure metals, the actinides, experimental ρ($T$) curves differ radically from one another to the point that there is no unified theoretical approach to understanding and fitting ρ($T$) data in these elements. First-principles calculations result in ρ($T$) curves that differ from experimental data, even qualitatively. In an attempt to unravel this long-standing problem, here I propose a simple model that accurately fits the ρ($T$) data for nine phases of elemental actinides (from thorium (Th) to curium (Cm)) for which experimental data are publicly available to date. The model is based on the concept of two parallel conduction channels: one is described by the Bloch-Grüneisen equation, which is associated with the classical electron-phonon dissipation mechanism, and the other by the Arrhenius equation, which is associated with the nearest-neighbor hopping (NNH) conductivity. Debye temperatures ($\Theta_D$) obtained by applying the model to the ρ($T$) data for nine elemental actinide phases are in good agreement with published values deduced from heat capacity measurements. For neptunium (Np) a maximum Arrhenius activation energy (among all actinides) of $E_a = 15.9$ meV was derived. The model was also successfully applied to ρ($T$) data measured on δ-phase plutonium-based alloys Pu-Ce and Pu-Ce-Ga.

*I. Introduction.* The earliest publicly available study of the $\rho(T)$ in uranium (U) which is one of the light actinide metals, dated by 1947 [1]. In the following decade, the $\rho(T)$ data for uranium alloys were published [2]. Sandenaw and Gibney [3] were the first to publish $\rho(T)$ data for the α-, β-, γ-, δ-, δ'-, and ε-phases of pure elemental plutonium (Pu) in 1958. Two years later, Sandenaw [4] published $\rho(T)$) data for the Pu-8 at.% Al alloy. And the following year, G. T. Meaden, in his Ph.D. thesis [5], published $\rho(T)$ data for elemental thorium (Th), uranium (U), neptunium (Np), α-plutonium (α-Pu), and the δ phase of Pu-Al alloys, which have since been reproduced in numerous reports [6–9]. In the following two decades, virtually the basic set of experimental $\rho(T)$ data for actinides and their alloys (publicly available to date) were published [6,7,10–28]. Some experimental $\rho(T)$ data for actinides alloys at ambient and high pressure were reported two decades ago [29,30]. Superconductivity and ferromagnetism in compounds and intermetallic compounds, where one metal is trans-uranium elements, are ongoing research topics [9,31–34]. Specific heat, thermal conductivity, magnetism, and elastic properties are other ongoing research topics [31,35–44] (and references therein) for actinides.

To demonstrate the challenges associated with the topic, two milestone plots are shown in Figure 1. The plot in Figure 1,a is from a study by Bett et al [7] (it had been published in 1984), while initially this type of plot was proposed by Hall et al [28] five decades ago in 1977.

Several models based on Matthiessen's rule were proposed to fit the experimental temperature-dependent resistivity $\rho(T)$ data measured in actinides. The primary equation within this approach is as the following [12,43,45–47]:

$$\rho(T) = \sum_{i=1}^{N} \rho_i(T), \quad (1)$$

where $\rho_i(T)$ represents a contribution from the *i*-th dissipation mechanism.

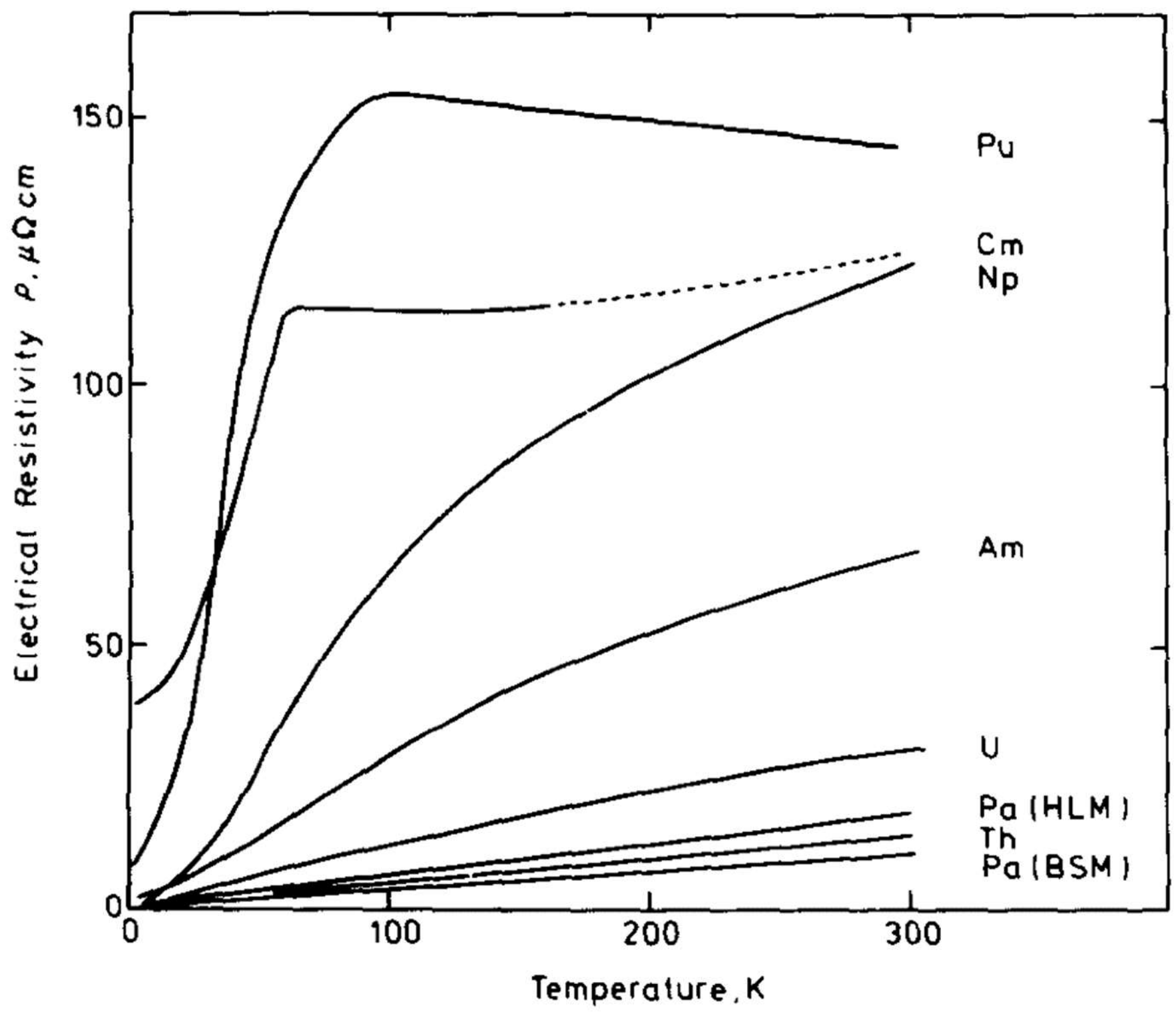

**Figure 1**. The temperature dependence of the electrical resistivities of the actinide metals. Plot is reproduced from Bett et al. [7] with permission from Elsevier (Licensed Content Publisher: Elsevier; Permission: License Number: 6315391153351; License date: Jul 24, 2026).

For instance, Schenkel [26] used the following equation for elemental curium (Cm):

$$\rho_{Cm}(T) = \rho_0 + \rho_{e-ph}(T) + \rho_{mag}(T), \tag{2}$$

where $\rho_0 \equiv \rho(T \to 0\text{ K})$ is the residual resistivity, $\rho_{e-ph}(T)$ is the contribution arising from the electron-phonon interaction (for which Schenkel [26] proposed to use the $\rho(T)$ dataset measured for pure americium (Am)), and $\rho_{mag}(T)$ is the contribution arising from magnetic interactions.

In some earlier versions of models based on Equation 2, total $\rho(T)$ is represented in polynomial form [24]:

$$\rho(T) = \sum_{i=1}^{N} \rho_i(T) = \sum_{i=1}^{N} A_i \times T^i \tag{3}$$

where, $i = 0{,}1{,}2{,}3{,}4$, and $A_i$ are free-fitting parameters. The main problem with this approach is that $A_i$ can significantly change and even change the sign at a minor change of experimental $\rho(T)$. Thus, the validity of free-fitting parameters $A_i$ derived from the fits by Equations 1,3 remain unclear.

Some models [7,48] are based on the power-law function:

$$\rho(T) = \rho_0 + A \times T^{\mathrm{n}}, \quad (4)$$

where $\rho_0$, $A$, and n are free-fitting parameters. The main problem with this approach is that a small change in experimental $\rho(T)$ causes a change in the value of n, and because of that, the units of free-fitting parameter $A$ are changing too. Based on that, a comparison of the derived parameter $A$, even for the samples made of the same actinide metal, is impossible.

More recently, Lawson [49] proposed a model to describe experimental $\rho(T)$ data measured in actinides alloys exhibited δ-Pu structure by a model based on four resistors connected in a particular serial-parallel configuration (Figure 3 in [49]).

First-principles calculations [50–53] were implemented to understand experimental $\rho(T)$ datasets for elemental actinides. These calculations were only particularly successful because the calculated $\rho(T)$ curves for Cm, α-Pu and β-Pu differed from experimental $\rho(T)$ curves even on a qualitative level (see also [54,55]).

For greater thoroughness, it should be mentioned that some $\rho(T)$ models (which were not applied to actinides) were developed for chemical compounds and multiphase samples [56–61], and these models are based on Matthiessen's rule (Equation 1):

$$\rho(T) = \rho_{BG}(T) + \rho_C(T), \quad (5)$$

where $\rho_0$ is the residual resistivity (which is a free fitting parameter), and

$$\rho_{\mathrm{BG}}(T) = \rho_0 + \rho_D \times \left(\frac{T}{\Theta_{\mathrm{D}}}\right)^5 \times \int_0^{\frac{\Theta_{\mathrm{D}}}{T}} \frac{x^5}{(e^x-1)(1-e^{-x})} dx, \quad (6)$$

is the Bloch-Grüneisen equation [62,63], where $\rho_D$ is the resistivity constant for the summation term associated with the Debye (acoustic) vibrations, and $\Theta_{\mathrm{D}}$ is the Debye temperature, and

$$\rho_C(T) = \rho_E \times \left(\frac{\Theta_\mathrm{E}}{T}\right) \times \frac{1}{\left(e^{\frac{\Theta_E}{T}}-1\right)\left(1-e^{-\frac{\Theta_E}{T}}\right)}, \quad (7)$$

is the Cooper equation [64], where $\rho_E$ is the resistivity constant for the summation term associated with the Einstein (optical) vibrations, and $\Theta_\mathrm{E}$ is the Einstein temperature.

*2. Proposed Model.* In an attempt to solve this long-standing problem, here I propose a simple universal model for $\rho(T)$ that accurately approximates all publicly available experimental $\rho(T)$ datasets measured in elemental actinides. The model is based on the understanding that the Matthiessen's rule (Equation 1) cannot be applied for single phase elemental metals, because the resistance arising from each dissipation mechanism should be associated with separate conduction channel, and, thus, total measured resistance is the reciprocal value for the sum of all conduction channels:

$$\rho(T) = \frac{1}{\sum_{i=1}^{N}\frac{1}{\rho_i(T)}}, \quad (8)$$

Experimenting with different equations for the $\rho_i(T)$ (in Equation 5), I found that the assumption of two conduction channels is sufficient to describe all reported $\rho(T)$ data for elemental actinides, as well as some data for actinides alloys. These channels are described by the following equations:

1. The Arrhenius equation [65]:

$$\rho_\mathrm{A}(T) = \rho_{sat} \times e^{\frac{E_a}{k_\mathrm{B}T}}, \quad (9)$$

where $\rho_\infty$ and $E_a$ are free-fitting parameters, and $\rho_{sat}$ is the resistivity constant in the Arrhenius part, $E_a$ is an activation energy, and $k_\mathrm{B}$ is the Boltzmann constant.

2. The Bloch-Grüneisen equation [62,63] which is given by Equation 6.

The explicit form of the proposed model is:

$$\rho(T) = \frac{1}{\sum_{i=1}^{2}\frac{1}{\rho_i(T)}} = \frac{1}{\frac{1}{\rho_{\mathrm{A}}(T)} + \frac{1}{\rho_{\mathrm{BG}}(T)}} = \frac{1}{\frac{1}{\rho_{sat}\times e^{\frac{E_a}{k_{\mathrm{B}}T}}} + \frac{1}{\rho_0 + \rho_D\times\left(\frac{T}{\Theta_{\mathrm{D}}}\right)^5\times\int_0^{\frac{\Theta_{\mathrm{D}}}{T}}\frac{x^5}{(e^x-1)(1-e^{-x})}dx}}, \tag{10}$$

To the best of my knowledge, this model has not been proposed earlier [8,9,13,65–78].

It should be noted that the model (Equation 10) has a well-known limit with $E_a = 0$, which is named the parallel resistor model [79,80]:

$$\left.\frac{1}{\rho(T)}\right|_{E_a=0} = \left.\frac{1}{\rho_{sat}\times e^{\frac{E_a}{k_{\mathrm{B}}T}}}\right|_{E_a=0} + \frac{1}{\rho_0 + \rho_D\times\left(\frac{T}{\Theta_{\mathrm{D}}}\right)^5\times\int_0^{\frac{\Theta_{\mathrm{D}}}{T}}\frac{x^5}{(e^x-1)(1-e^{-x})}dx} = \frac{1}{\rho_{sat}} + \frac{1}{\rho_0 + \rho_D\times\left(\frac{T}{\Theta_{\mathrm{D}}}\right)^5\times\int_0^{\frac{\Theta_{\mathrm{D}}}{T}}\frac{x^5}{(e^x-1)(1-e^{-x})}dx}, \tag{11}$$

where the designation of $\rho_{sat}$ is taken from the terminology of the parallel resistance model [79,80].

The parallel resistor model (Equation 11) [79,80] is widely used for a variety of conductors [81–89] since its introduction in 1977. The physics behind this model (including the Ioffe-Rigel criterion [90,91]) had been discussed in great detail [8,66,68,78,92–99].

In a way, Equation 10 can be considered a generalization of the parallel resistance model (Equation 11 [79,80]) by understanding that the saturated resistance part can exhibit temperature dependence described by the Arrhenius law:

$$\frac{1}{\rho_{sat}} \to \frac{1}{\rho_{sat}\times e^{\frac{E_a}{k_{\mathrm{B}}T}}}. \tag{12}$$

Hypothetically, for some conductors, the saturated resistance term can be further advanced by the Mott-type [73] or the Efros-Shklovskii-type temperature dependence [75]:

$$\frac{1}{\rho_{sat}} \to \frac{1}{\rho_{sat}\times e^{\left(\frac{E_a}{k_{\mathrm{B}}T}\right)^p}}, \tag{13}$$

where $p = \frac{1}{1+N}$, and $1 \le N \le 3$. In the result, the full equation for this case is:

$$\rho(T) = \frac{1}{\frac{1}{\rho_{sat} \times e^{\left(\frac{E_a}{k_B T}\right)^p}} + \frac{1}{\rho_0 + \rho_D \times \left(\frac{T}{\Theta_D}\right)^5 \times \int_0^{\frac{\Theta_D}{T}} \frac{x^5}{(e^x - 1)(1 - e^{-x})} dx}}, \quad (14)$$

However, to date, I have not found any experimental $\rho(T)$ data that requires the use of a more complicated model described by Equation 14, instead of Equation 10.

Based on the above, it seems reasonable to call the model described by Equation 10 the nearest-neighbor hopping parallel resistance model (NNH-parallel resistance model). This model has an additional free-fitting parameter, $E_a$, compared to the well-known parallel resistance model [79,80,91].

Similarly, the model described by Equation 14 is reasonable to call the variable-range hopping parallel resistance model (VRH- parallel resistance model). This model has two additional free-fitting parameters ($E_a$ and $p$) compared to the well-known parallel resistance model [79,80,91].

For greater thoroughness, it should be noted that the exponent of the power law in the Bloch-Grüneisen equation [62,63] (Equation 6) can also be a freely adjustable parameter [100]. Based on that most general equation describing the hopping parallel resistance model can be written in the following form:

$$\rho(T) = \frac{1}{\frac{1}{\rho_{sat} \times e^{\left(\frac{E_a}{k_B T}\right)^p}} + \frac{1}{\rho_0 + \rho_D \times \left(\frac{T}{\Theta_D}\right)^m \times \int_0^{\frac{\Theta_D}{T}} \frac{x^m}{(e^x - 1)(1 - e^{-x})} dx}}, \quad (15)$$

where $\rho_{sat}$, $E_a$, $p$, $\rho_0$, $\rho_D$, $\Theta_D$, and $m$ are free-fitting parameters.

The results of the analysis of the $\rho(T)$ datasets for pure elemental actinides (as they appear in the Mendeleev table (from thorium to curium)) using the NNH-parallel resistance model (Equation 10) are presented below.

*3. Results and Discussion. 3.1. Thorium.* Thorium (Th) is the lightest actinide metal for which experimental $\rho(T)$ data have been measured. Formally, it is the second element in the actinide group of the Mendeleev table, after actinium (Ac). However, because all isotopes of actinium have very short half-lives, experimental $\rho(T)$ data are not available for pure actinium.

Schettler et al [25] reported $\rho(T)$ datasets for three thorium samples, which were designated as Sample 1, Sample 2, and Sample 3. Figure 2 shows the fits of the $\rho(T)$ datasets for these three samples with Equation 10.

One can see that fits eliminate the parallel resistor term, $\frac{1}{\rho_{sat}\times e^{\frac{E_a}{k_B T}}}$, and standard Bloch-Grüneisen equation [62,63] (Equation 6) perfectly approximates raw experimental data for three thorium samples (Figure 2). Deduced Debye temperature $143\ \mathrm{K} \leq \Theta_\mathrm{D} \leq 177\ \mathrm{K}$ is in the same ballpark as independently reported values of $\Theta_\mathrm{D} = 160\ \mathrm{K}$ [101], $\Theta_\mathrm{D} = 163\ \mathrm{K}$ [101], $\Theta_\mathrm{D} = 144\ \mathrm{K}$ [7].

Besides the above, Figure 3 presents result of the global fit of the $\rho(T)$ data for these three samples. The fit was performed with the following system of equations:

$$\begin{cases} \rho_{Sampl}\quad(T) = \dfrac{1}{\frac{1}{\rho_{sat,global}\times e^{\frac{E_{a,global}}{k_B T}}} + \frac{1}{\rho_{0,sample1}+\rho_{D,Sample1}\times\left(\frac{T}{\Theta_{\mathrm{D,global}}}\right)^5\times\int_0^{\frac{\Theta_{\mathrm{D,global}}}{T}}\frac{x^5}{(e^x-1)(1-e^{-x})}dx}} \\ \rho_{Sample2}(T) = \dfrac{1}{\frac{1}{\rho_{sat,global}\times e^{\frac{E_{a,global}}{k_B T}}} + \frac{1}{\rho_{0,sample2}+\rho_{D,Sample2}\times\left(\frac{T}{\Theta_{\mathrm{D,global}}}\right)^5\times\int_0^{\frac{\Theta_{\mathrm{D,global}}}{T}}\frac{x^5}{(e^x-1)(1-e^{-x})}dx}} \\ \rho_{Sample3}(T) = \dfrac{1}{\frac{1}{\rho_{sat,global}\times e^{\frac{E_{a,globa}}{k_B T}}} + \frac{1}{\rho_{0,sample3}+\rho_{D,Sample3}\times\left(\frac{T}{\Theta_{\mathrm{D,global}}}\right)^5\times\int_0^{\frac{\Theta_{\mathrm{D,global}}}{T}}\frac{x^5}{(e^x-1)(1-e^{-x})}dx}} \end{cases}, \tag{16}$$

where $\rho_{sat,global}$, $E_{a,global}$, and $\Theta_{\mathrm{D,global}}$ are free-fitting global parameters, and $\rho_{0,sample1}$, $\rho_{D,Sa}$ , $\rho_{0,sample2}$, $\rho_{D,Sample2}$, $\rho_{0,sample3}$, $\rho_{D,Sampl}$ are free-fitting local parameters.

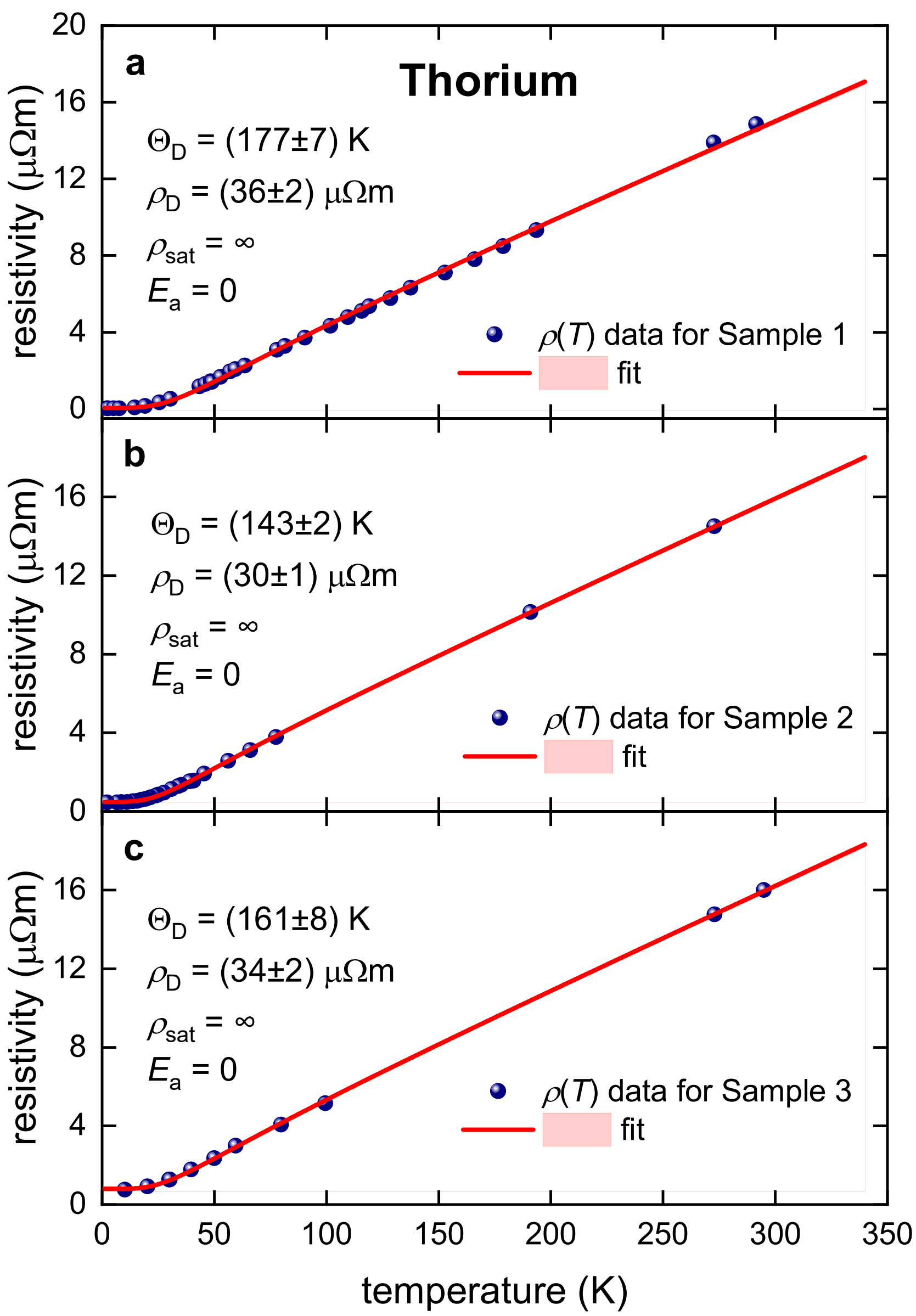

**Figure 2**. Fits with the nearest-neighbor hopping parallel resistance model (Equation 10) of three $\rho(T)$ datasets for pure thorium samples reported by Schettler et al [25]. (**a**) Sample 1 has $\frac{\rho(T=273\text{ K})}{\rho(T=4.2\text{ K})} = 480$. Fit quality $R$-square (COD) = 0.9989. (**b**) Sample 2 has $\frac{\rho(T=273\text{ K})}{\rho(T=4.2\text{ K})} = 140$. Fit quality $R$ = 0.9998. (**c**) Sample 3 has $\frac{\rho(T=2\quad\text{K})}{\rho(T=4.2\text{ K})} = 31$. Fit quality $R$ = 0.9998. Deduced parameters are shown in each panel. The 95% confidence bands are shown by pink shadow areas.

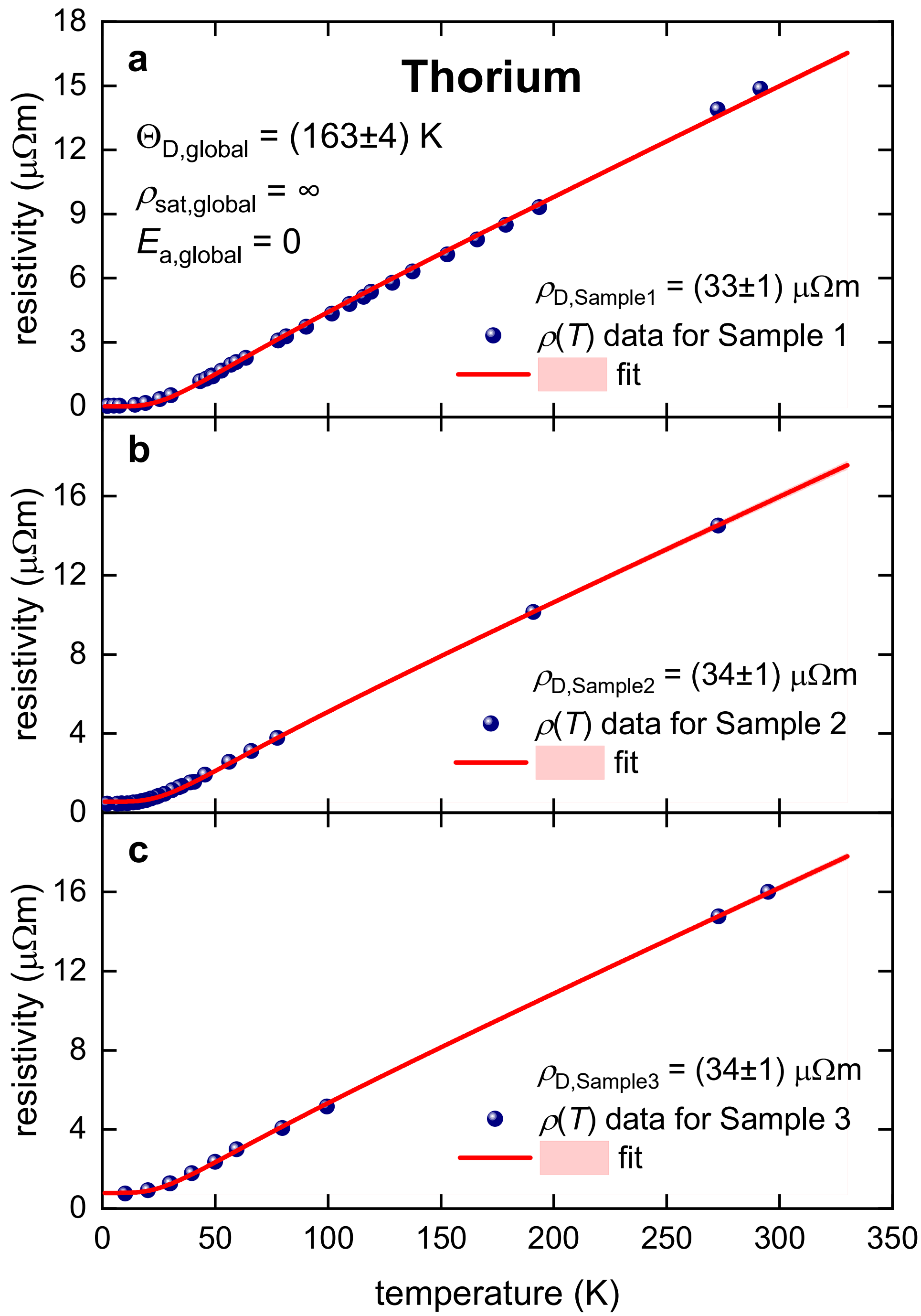

**Figure 3**. Global fits with the nearest-neighbor hopping parallel resistance model (Equation 16) of three $\rho(T)$ datasets for pure thorium samples reported by Schettler et al [25]. Global fit quality $R$-square (COD) = 0.9993. (**a**) Sample 1 has $\frac{\rho(T=273\text{ K})}{\rho(T=4.2\text{ K})} = 480$. (**b**) Sample 2 has $\frac{\rho(T=273\text{ K})}{\rho(T=4.2\text{ K})} = 140$. (**c**) Sample 3 has $\frac{\rho(T=273\text{ K})}{\rho(T=4.2\text{ K})} = 31$. Deduced parameters are shown in the panels. The 95% confidence bands are shown by pink shadow areas.

Figure 3 demonstrates that the global data fit shows that the independent free-fit parameters $\rho_{D,i}$ are in good agreement with each other, and the deduced Debye temperature $\Theta_\mathrm{D} = (163 \pm 4)$ K is the same as independently reported $\Theta_\mathrm{D}$ values (see references [7,101,102]) for thorium.

*3.2. Protactinium.* Protactinium (Pa) is the third element in the actinides group in the Mendeleev table. Raw experimental $\rho(T)$ datasets for pure protactinium were reported by Bett et al [7] and by Hall et al [28]. Reported $\rho(T)$ datasets for these samples together with the fits with Equation 10 are shown in Figure 4.

Similarly, to the case of thorium, fits eliminate the parallel resistor term, $\frac{1}{\rho_{sat} \times e^{\frac{E_a}{k_\mathrm{B} T}}}$, and standard Bloch-Grüneisen equation [62,63] (Equation 6) perfectly approximates raw experimental $\rho(T)$ data for both samples (Figure 4). Deduced Debye temperature $\Theta_\mathrm{D} = (106 \pm 4)$ K for the sample reported by Hall et al [28] is in good agreement with the value reported in the original paper, $\Theta_\mathrm{D} = 108$ K [28]. Other research teams reported a higher Debye temperature. For instance, Bett et al [7] reported $\Theta_\mathrm{D} = 147$ K; Stewart et al [103] reported $\Theta_\mathrm{D} = (185 \pm 5)$ K; and Mortimer [104] reported $\Theta_\mathrm{D} = 180$ K.

It is worth noting that the last two values, $\Theta_\mathrm{D} = (185 \pm 5)$ K [103] and $\Theta_\mathrm{D} = 180$ K [104], were deduced from heat capacity measurements, and these values are the same as the value deduced in Figure 4,b, $\Theta_\mathrm{D} = (178 \pm 3)$ K, from the $\rho(T)$ data reported by Bett et al [7].

It was mentioned by Bett et al [7] that "*…there are considerable differences*" between experimental data reported in two studies [7,28]. The discussion can be found in Ref. [7]. Important difference mentioned by Bett et al [7] is that their Pa samples exhibited the lowest absolute $\rho(T)$ values across all actinides (which is more than twice lower than Pa samples studied by Hall et. al. [28]). The likely origin is that the samples studied by Bett et al. [7] were extremely purified (with a total impurity content of 0.05 at.%).

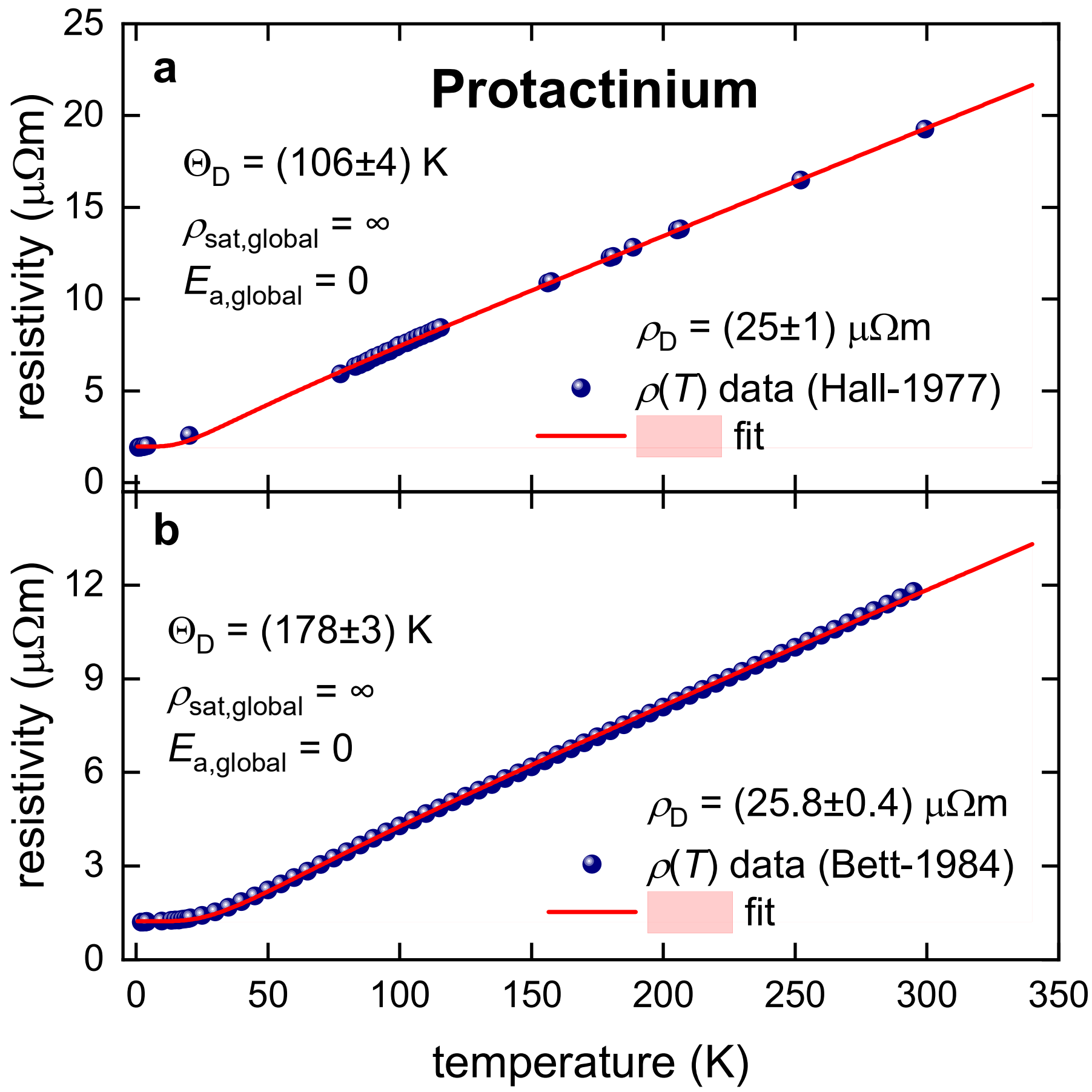

**Figure 4**. Fits with the nearest-neighbor hopping parallel resistance model (Equation 10) of two $\rho(T)$ datasets for pure protoactinium samples reported by (**a**) Hall et al [28], and by (**b**) Bett et al [7]. (**a**) Fit quality $R$-square (COD) = 0.9997. (**b**) Fit quality $R$-square (COD) = 0.9998. Deduced parameters are shown in each panel. The 95% confidence bands are shown by pink shadow areas.

In addition, the raw experimental data for $\rho(T)$ reported by Hall et al. [28] are very non-uniformly distributed across the entire temperature range (Figure 4,a), which is not a preferable option for any comprehensive data analysis.

Therefore, it can be concluded that the analysis of the $\rho(T)$ data reported by Bett et al. [7] using Equations 6 and 10 (in Figure 4b) independently confirms that the Debye temperature of pure elemental protactinium is $\Theta_D = 180$ K, a value first reported almost five decades ago [103,104].

*3.3. Uranium.* This actinide metal is one of the primary chemical elements in modern energy technology. Under ambient conditions, uranium exists in the α-phase, and at a temperature of approximately T≅40 K, it transforms into the α′-phase [105]. This transition is visible in the $\rho(T)$ curves as an inflection point [5,11,12,23], which is highlighted in Figure 5 by arrows.

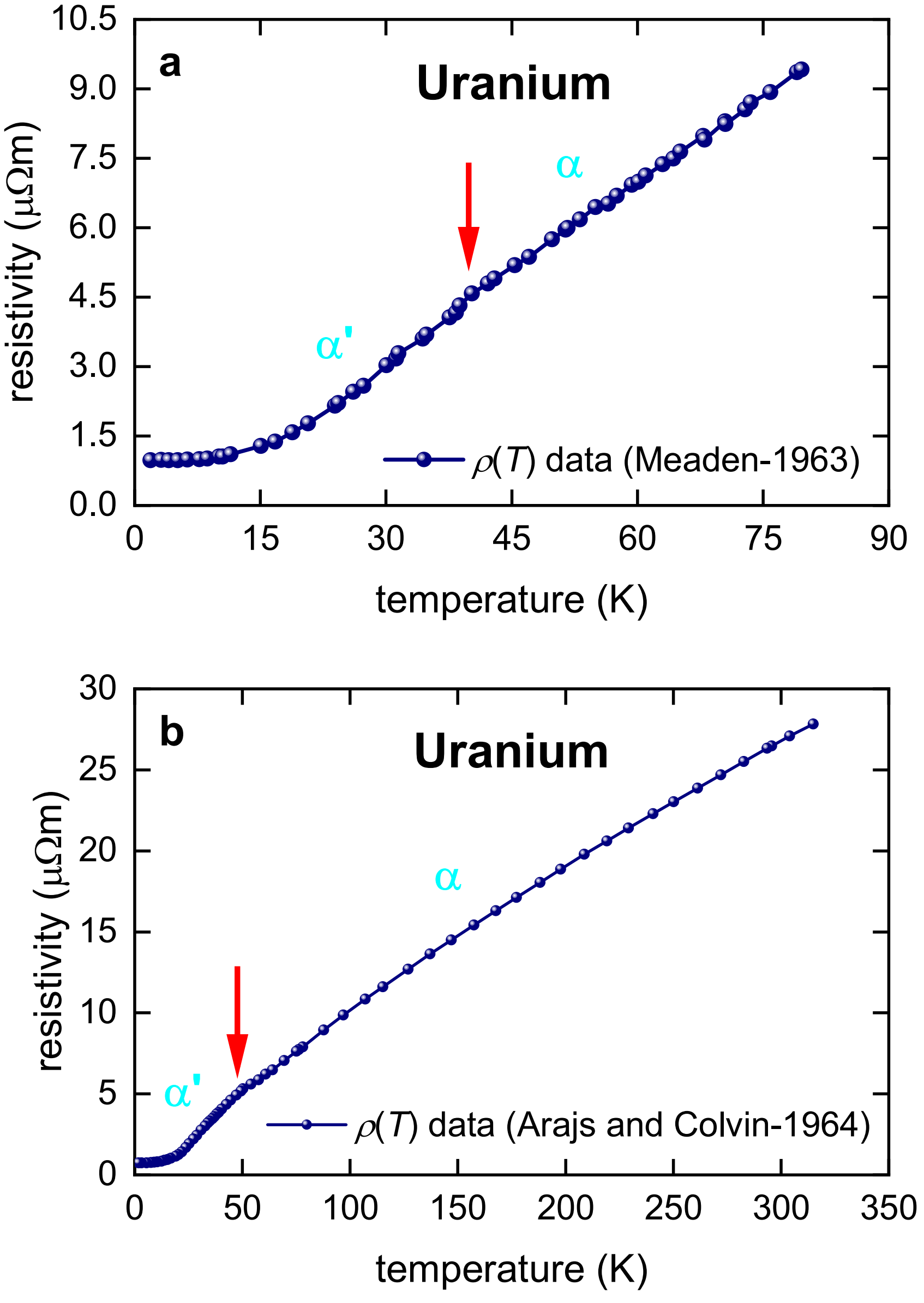

**Figure 5**. $\rho(T)$ data for pure uranium samples reported by (**a**) Meaden [11], and by (**b**) Arajs and Colvin (Sample U-1) [23]. Red arrows show the inflection point for the α ⇆ α′ phase transition.

On that basis, the fitting of the experimental data $\rho(T)$ for the U-1 uranium sample reported by Arajs and Colvin [23] with Equation 10 has been performed separately for the α′ and α phases in Figure 6.

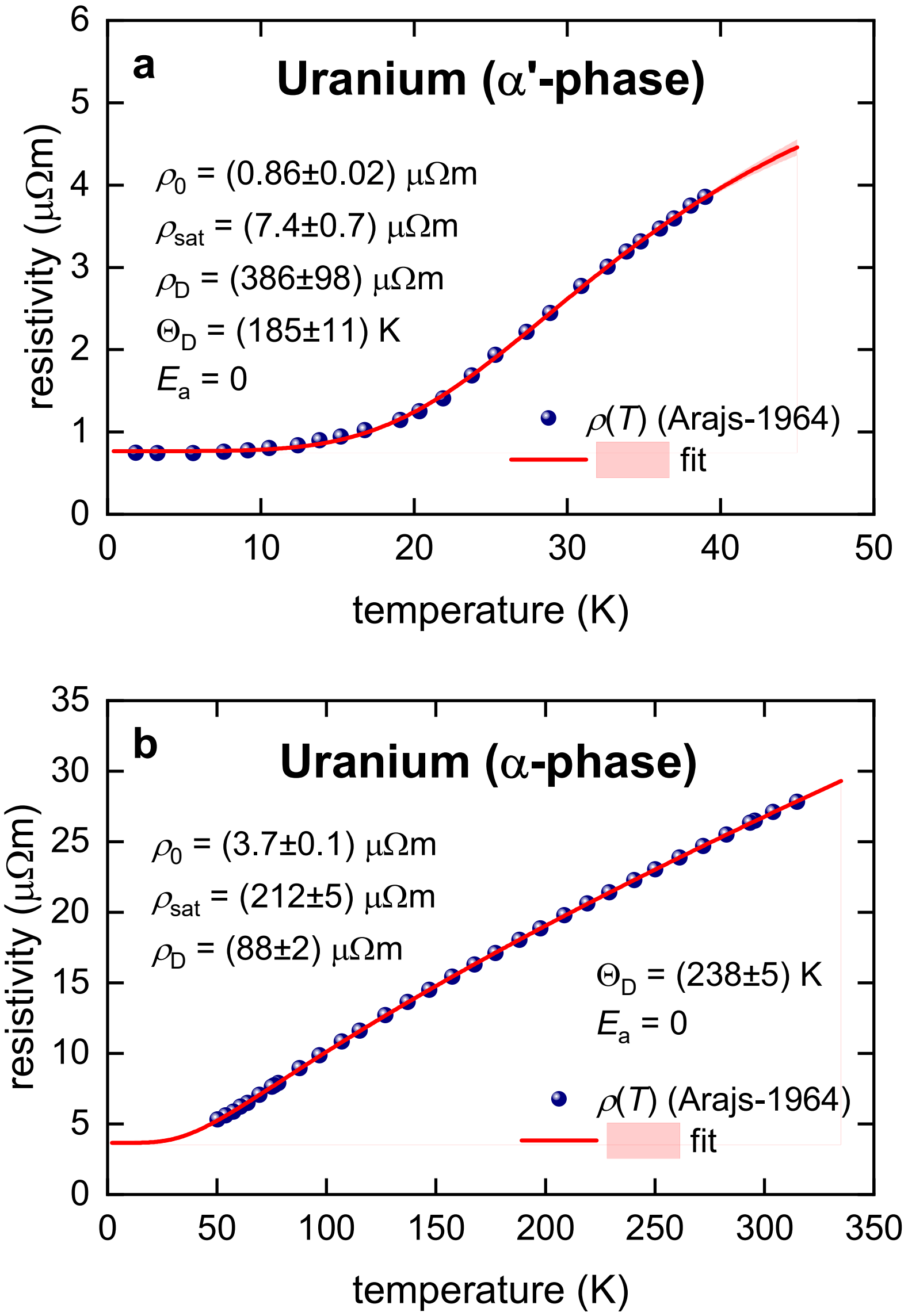

**Figure 6**. Fits with the nearest-neighbor hopping parallel resistance model (Equation 10) of the $\rho(T)$data for the pure uranium Sample U-1 reported by Arajs and Colvin [23]. (**a**) low-temperature α′-phase of uranium. Fit quality *R*-square (COD) = 0.9996. (**b**) α-phase of uranium. Fit quality *R* = 0.9999. Deduced parameters are shown in each panel. The 95% confidence bands are shown by pink shadow areas.

The deduced Debye temperature $\Theta_D = (185 \pm 11)$ K for the α′-phase (Figure 6,a) is in good agreement with the value of $\Theta_D(T = 35\text{ K}) = 169$ K [106], $\Theta_D = 162$ K [107], and $\Theta_{\mathrm{D}}(T=0\ \mathrm{K})=170\ \mathrm{K}$ [12] (three latest values were deduced from low-temperature heat-capacity measurements).

The deduced Debye temperature $\Theta_{\mathrm{D}}=(238\pm5)$ K for the α-phase (Figure 6,b) is in good agreement with the value of $\Theta_{\mathrm{D}}=248\ \mathrm{K}$ [101], and $\Theta_{\mathrm{D}}=222\ \mathrm{K}$ [108].

It is also important to note that the fits show that the saturated resistance term, $\frac{1}{\rho_{sat}\times e^{\frac{E_a}{k_{\mathrm{B}}T}}}\rightarrow\frac{1}{\rho_{sat}}$, is required to approximate $\rho(T)$ data for both phases of uranium Sample U-1 (and free-fitting value for the activation energy is zero, $E_a=0$).

*3.4. Neptunium.* $\rho(T)$ data for a pure neptunium sample reported by Olsen and Elliot [16] is shown in Figure 7 together with the fit with the NNH-parallel resistor model (Equation 10).

One of the most interesting results of the fit (Figure 7) is that the saturated resistance term, $\frac{1}{\rho_{sat}\times e^{\frac{E_a}{k_{\mathrm{B}}T}}}$, is necessary to approximate the $\rho(T)$ data, and the deduced activation energy is not zero, but is $E_a=(15.9\pm0.6)$ meV. This energy is equivalent to the temperature $\frac{E_a}{k_{\mathrm{B}}}=(185\pm7)$ K.

Deduced Debye temperature $\Theta_{\mathrm{D}}=187$ K is in the range reported by other research teams: $\Theta_{\mathrm{D}}=150\ \mathrm{K}$ [23], and $\Theta_{\mathrm{D}}=163\ \mathrm{K}$ [109].

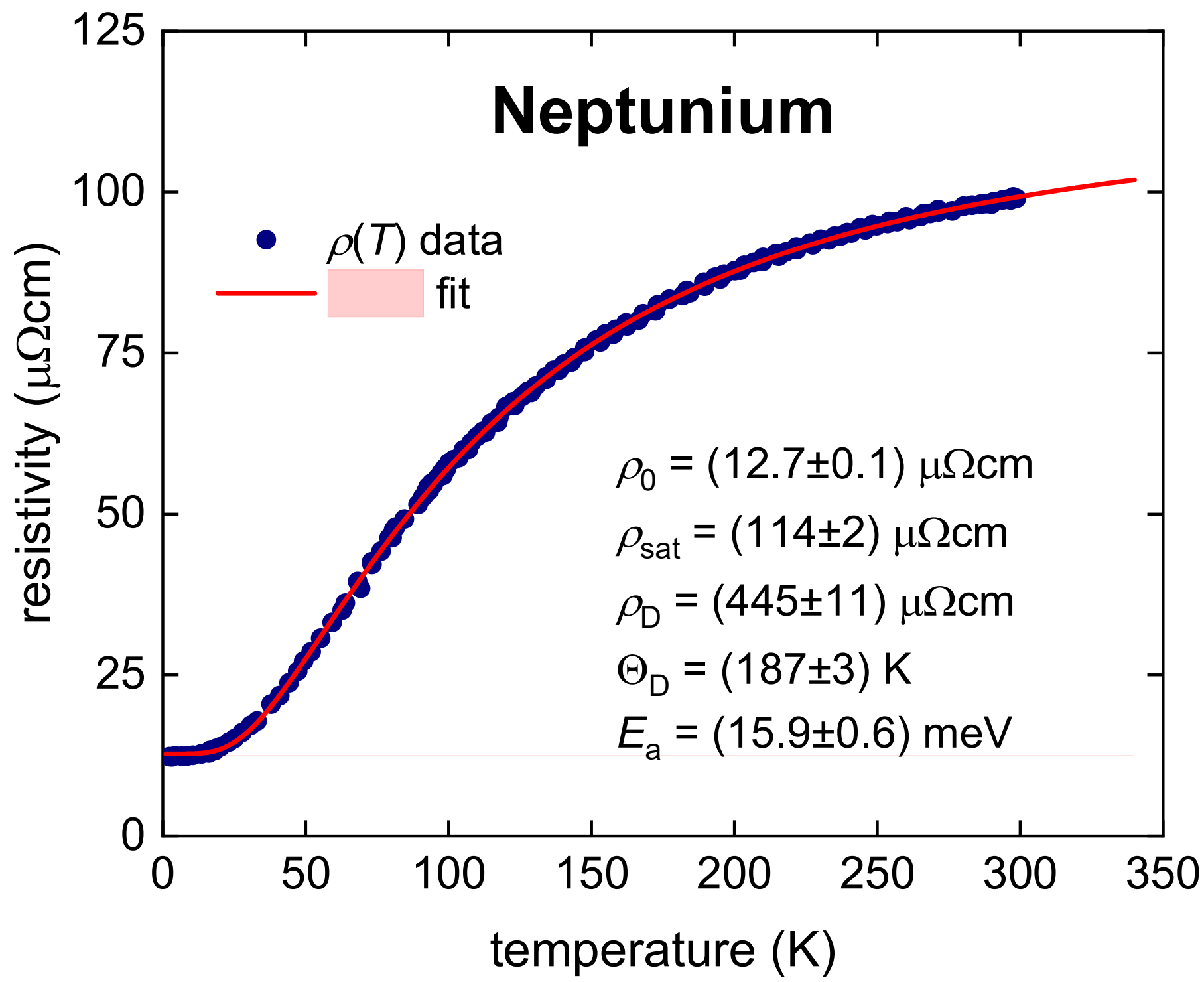

**Figure 7**. The fitting of the nearest-neighbour hopping parallel resistance model (Equation 10) to the $\rho(T)$data for the pure neptunium sample reported by Olsen and Elliot [16]. Fit quality $R$-square (COD) = 0.9998. Deduced activation Arrhenius energy, $E_a = (15.9 \pm 0.6)$ meV, is equivalent to the temperature $\frac{E_a}{k_B} = (185 \pm 7)$ K. Other deduced parameters are shown in the figure. The 95% confidence bands are shown by pink shadow areas.

*3.5. α-Plutonium.* Figure 8 shows $\rho(T)$ datasets for a pure α-plutonium samples reported by King and Lee [110] (Figure 8,a), Olsen and Elliot [16] (Figure 8,b), and Meaden [5,11] together with the fits with the NNH-parallel resistance model (Equation 10).

Remarkably, to note two findings in Figure 8:

3.5.1. Saturated resistance constant $\rho_{sat} \cong 135$ $\mu\Omega$cm is practically the same for all three samples, despite these different samples were studied by three independent research groups.

3.5.2. Arrhenius activation energy $E_a = (4.6 \pm 0.2)$ meV is the same for samples studied by King and Lee [110] (Figure 8,a) and Meaden [5,11,12]. The value of $E_a = (5.8 \pm 0.1)$ meV derived for the $\rho(T)$ data reported Olsen and Elliot [16] is in the same ballpark.

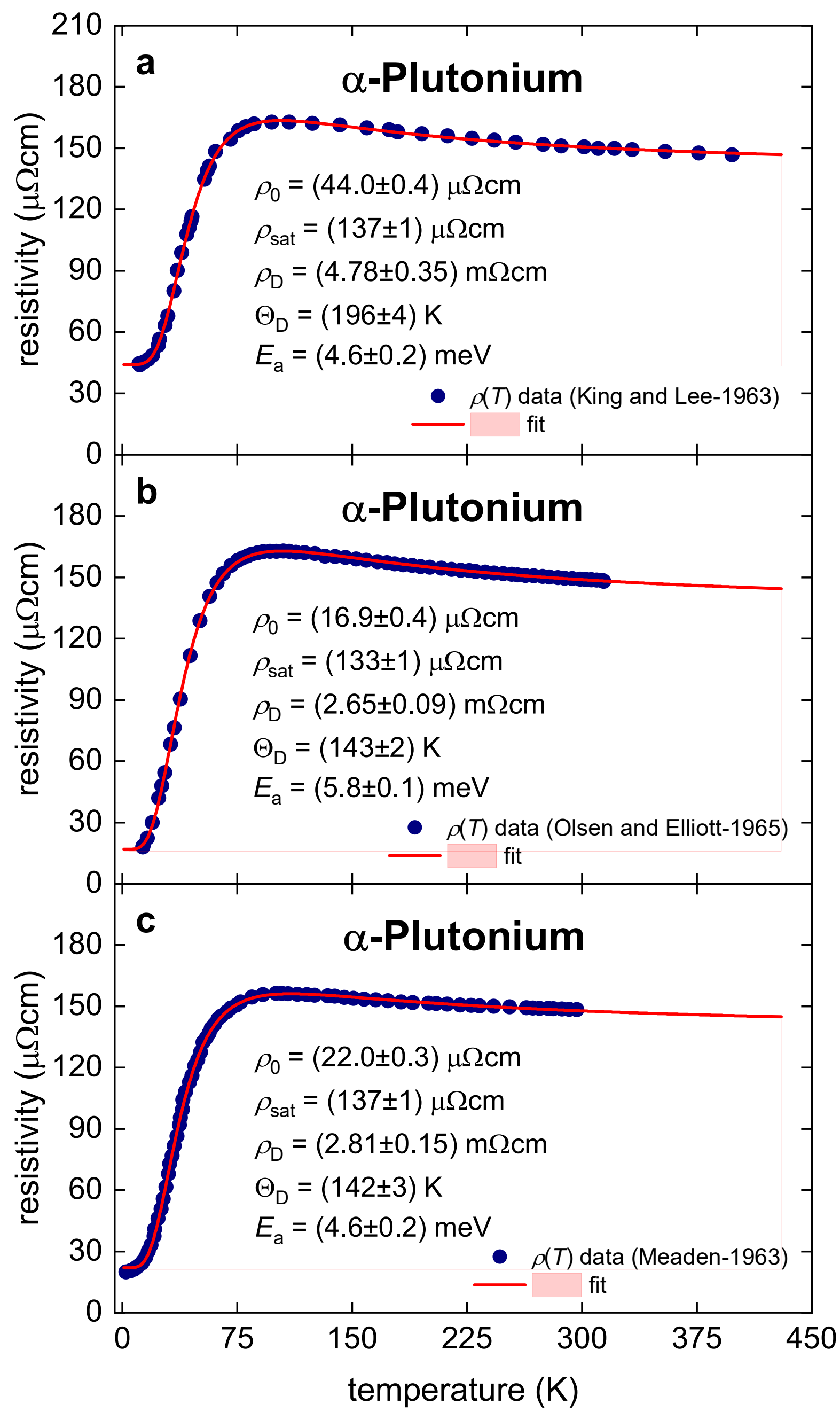

**Figure 8**. The fitting of the nearest-neighbor hopping parallel resistance model (Equation 10) of the $\rho(T)$ data for α-plutonium samples. (**a**) $\rho(T)$ data reported by King and Lee [110] (dataset digitized from Figure 17 in Reference [12]). Deduced activation Arrhenius energy, $E_a = (4.6 \pm 0.2)$ meV, is equivalent to the temperature $\frac{E_a}{k_B} = (54 \pm 2)$ K. Fit quality $R$-square (COD) = 0.9997. (**b**) $\rho(T)$ data reported by Olsen and Elliot [16]. Deduced activation Arrhenius energy, $E_a = (5.8 \pm 0.1)$ meV, is equivalent to the temperature $\frac{E_a}{k_B} = (67 \pm 1)$ K. Fit quality 0.9998. (**c**) $\rho(T)$ data reported by Meaden [5,11,12]. Deduced activation Arrhenius energy, $E_a = (4.6 \pm 0.2)$ meV, is equivalent to the temperature $\frac{E_a}{k_B} = (54 \pm 2)$ K. Fit quality 0.9996. Other deduced parameters are shown in the figure. The 95% confidence bands are shown by pink shadow areas.

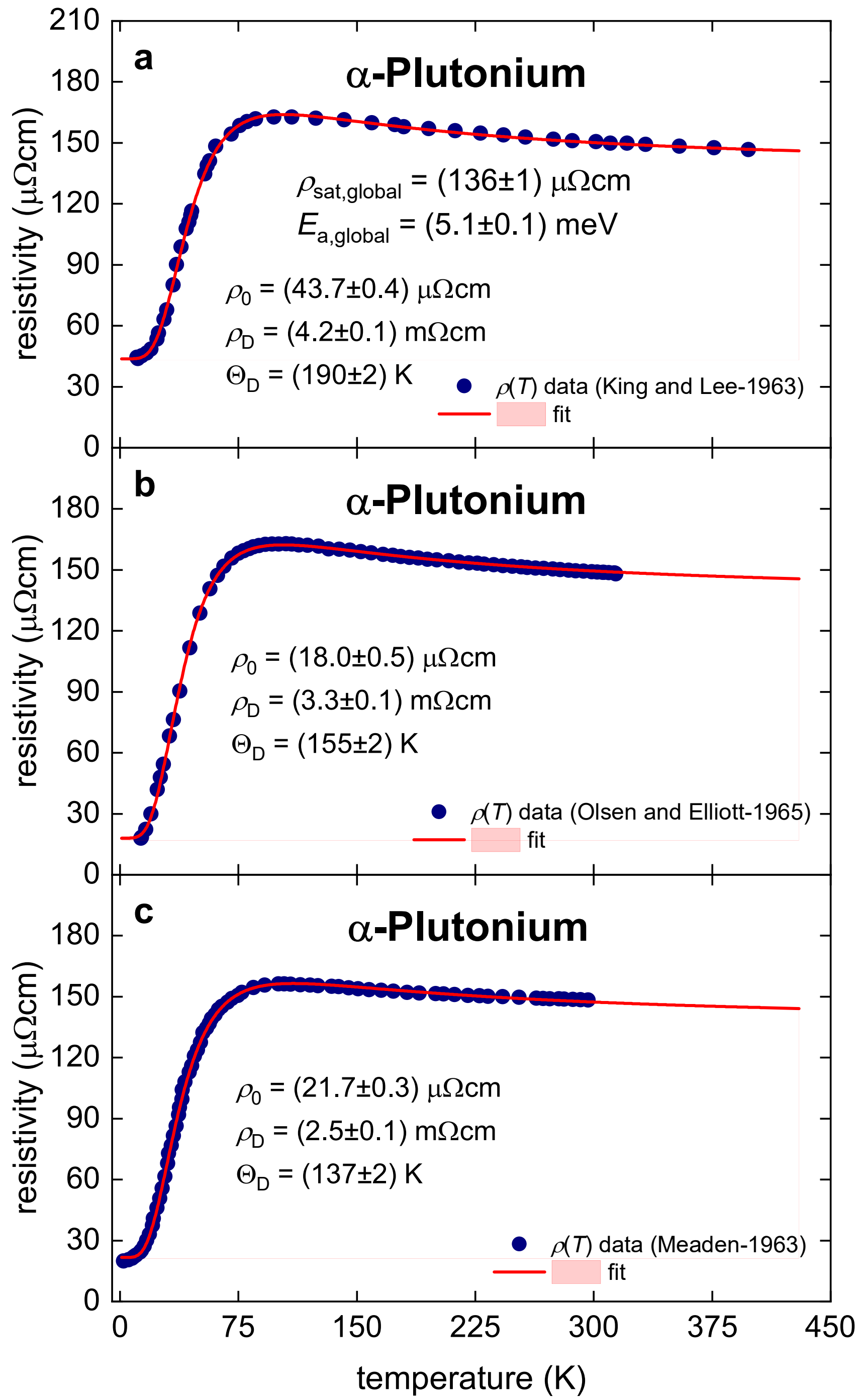

**Figure 9**. Global fit with the nearest-neighbor hopping parallel resistance model (Equation 10) of the $\rho(T)$ datasets for the pure α-plutonium samples. Deduced global parameters are: $E_{\text{a,global}} = (5.1 \pm 0.1)$ meV, and the equivalent temperature $\frac{E_{\text{a,global}}}{k_B} = (59 \pm 1)$ K, and $\rho_{\text{sat,global}} = (136 \pm 1)$ μΩcm. Fit quality *R*-square (COD) = 0.9996. (**a**) $\rho(T)$ data reported by King and Lee [110] (dataset digitized from Figure 17 in Reference [12]). (**b**) $\rho(T)$ data reported by Olsen and Elliot [16]. (**c**) $\rho(T)$ data reported by Meaden [5,11,12]. Other deduced parameters are shown in the figure. The 95% confidence bands are shown by pink shadow areas.

To derive the average values for these parameters, the global data fit (by Equation 16) has been performed, and it is shown in Figure 9. The deduced values are:

3.5.3. Saturated resistance constant $\rho_{\text{sat,global}} = (136 \pm 1)\ \mu\Omega\text{cm}$.

3.5.4. Arrhenius activation energy $E_{\text{a,global}} = (5.1 \pm 0.1)$ meV, and the equivalent Arrhenius temperature is $\frac{E_{\text{a,global}}}{k_{\text{B}}} = (59 \pm 1)$ K.

Deduced Debye temperature $137\ \text{K} \leq \Theta_{\text{D}} \leq 190\ \text{K}$ is in the overall range reported by other research teams: $\Theta_{\text{D}} = 175\ \text{K}$ [12], and $\Theta_{\text{D}} = 206\ \text{K}$ [101], $\Theta_{\text{D}} = 153\ \text{K}$ [36], and $\Theta_{\text{D}} = 116\ \text{K}$ [111], $\Theta_{\text{D}} = 175\ \text{K}$ [31], and values reported at high temperatures $\Theta_{\text{D}}(T = 380\ \text{K}) = 207\ \text{K}$ [31] and $\Theta_{\text{D}}(T = 380\ \text{K}) = 205\ \text{K}$ [31].

*3.6. β-Plutonium.* Figure 10 shows the $\rho(T)$ dataset for high-purity β-plutonium sample (reported by King and Lee [110] (in their Figure 2 [110])) and data fit with the nearest-neighbour hopping parallel resistance model (Equation 10). Fit has a high quality despite a very limited number of measured $\rho(T)$ data points in the database.

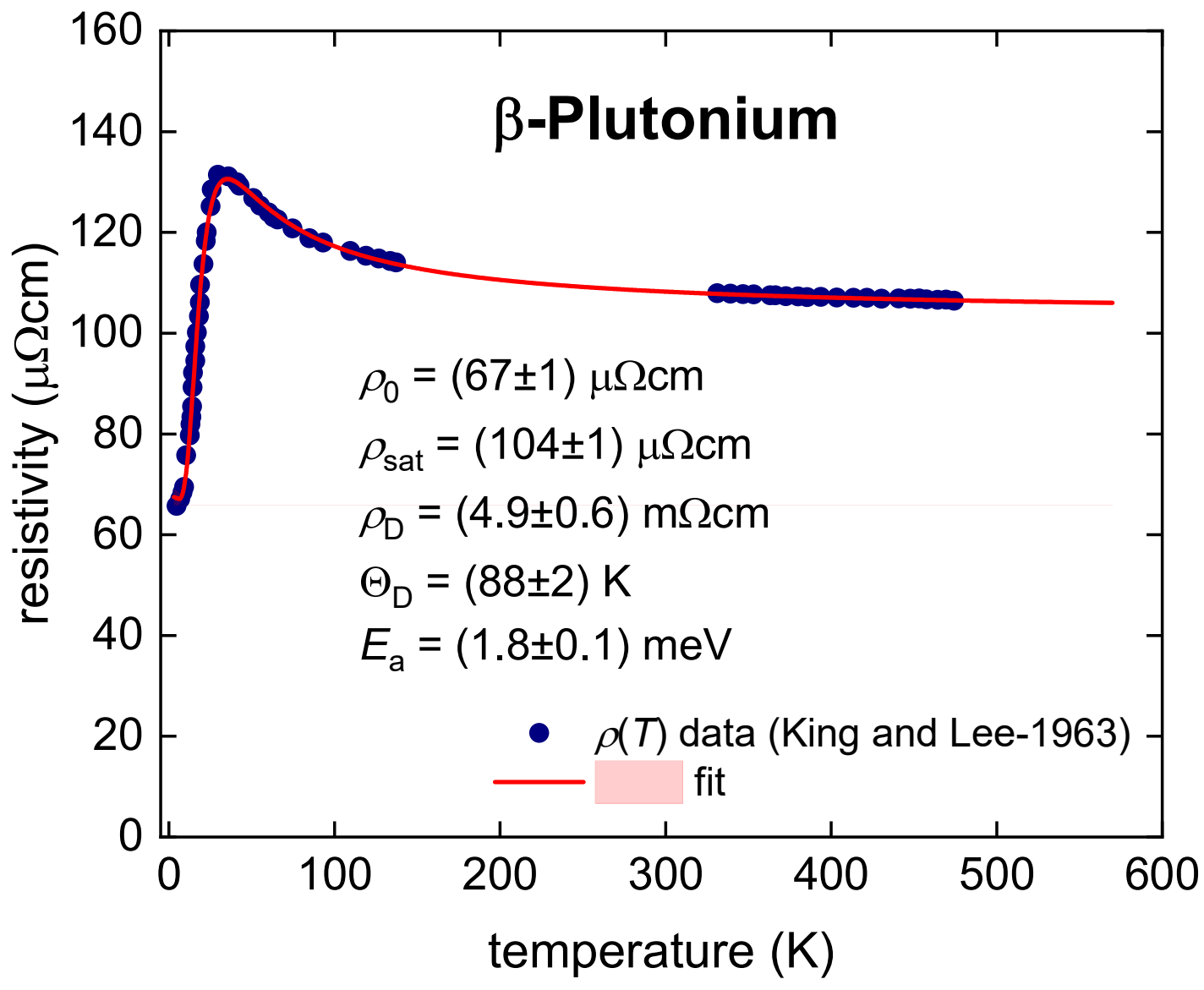

**Figure 10**. The fitting of the nearest-neighbor hopping parallel resistance model (Equation 10) of the $\rho(T)$ data for high-purity β-plutonium sample reported by King and Lee [110]. Deduced activation

Arrhenius energy, $E_a = (1.8 \pm 0.1)$ meV, is equivalent to the temperature $\frac{E_a}{k_B} = (21 \pm 1)$ K. Fit quality $R$-square (COD) = 0.9962. Other derived parameters are shown in the figure. The 95% confidence bands are shown by pink shadow areas.

There are independently reported values for the Debye temperature in β-plutonium $\Theta_D = 100$ K [36], $\Theta_D = 102$ K [40], and $\Theta_D = 71$ K [111]. Derived Arrhenius activation energy $E_a = (1.8 \pm 0.1)$ meV has an equivalent temperature scale of $\frac{E_a}{k_B} = (21 \pm 1)$ K.

*3.7. Americium.* Figure 11 shows the $\rho(T)$ datasets for pure americium samples reported by Schenkel and Müller [27], and Müller et al. [22] and data fit with the nearest-neighbour hopping parallel resistance model (Equation 10). Figure 12 shows the same $\rho(T)$ datasets and global data fit with the nearest-neighbour hopping parallel resistance model (Equation 16).

Deduced global Debye temperature $\Theta_{D,global} = (110 \pm 2)$ K is in a good agreement with the values reported by Griveau and Colineau, $\Theta_D = 121$ K [108], Smith et al., $\Theta_D = 121$ K [44], and Bett et al., $\Theta_D = 110$ K [7], Suzuki et al. $\Theta_D = 116$ K [31].

It should be noted that the Arrhenius energy is zero for all americium samples $E_{a,global} = 0$, while $\rho_{sat} \neq 0$ and therefore the data were fitted by the saturated resistance model [79].

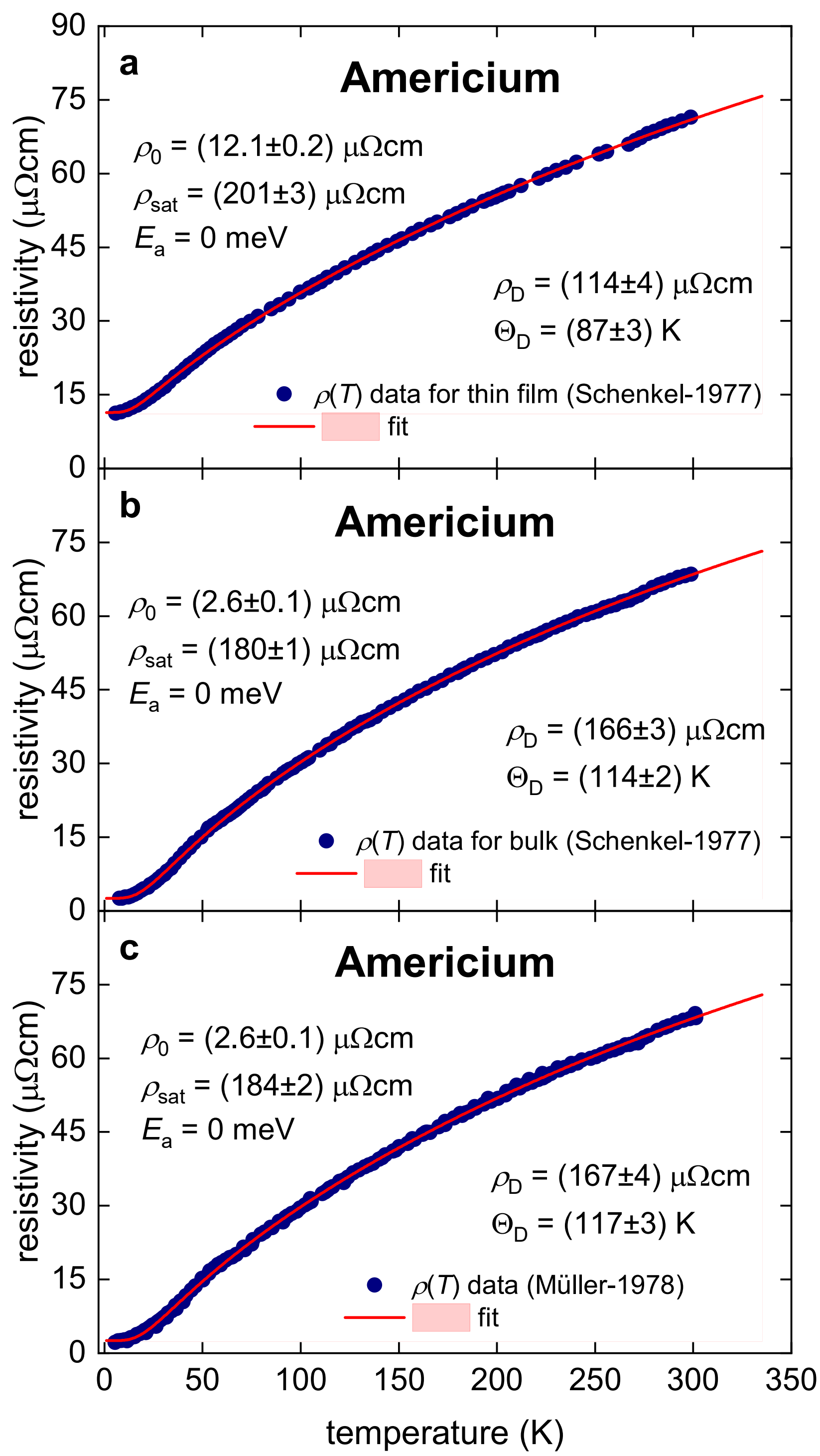

**Figure 11**. The fitting by the nearest-neighbor hopping parallel resistance model (Equation 10) of the $\rho(T)$ data for α-americium samples. (**a**) $\rho(T)$ data reported for thin film sample by Schenkel and Müller [27]. Fit quality *R*-square (COD) = 0.9998. (**b**) $\rho(T)$ data reported for bulk sample sample by Schenkel and Müller [27]. Fit quality 0.9999. (**c**) $\rho(T)$ data reported for thin film sample by Müller et al [22]. Fit quality 0.9997. Other deduced parameters are shown in the figure. The 95% confidence bands are shown by pink shadow areas.

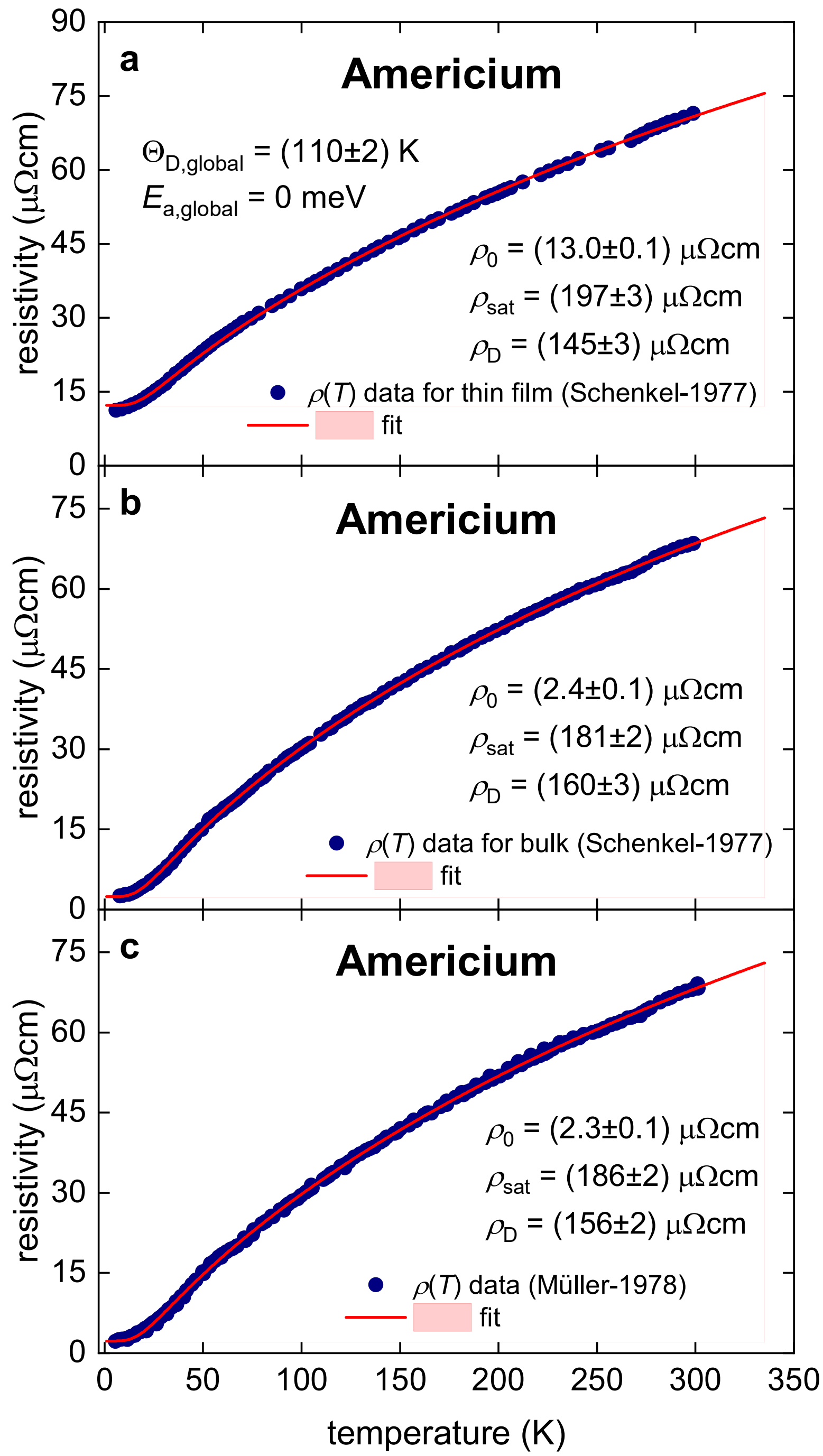

**Figure 12**. The global fitting by the nearest-neighbor hopping parallel resistance model (Equation 10) of the $\rho(T)$ data for α-americium samples. Deduced global parameters: Debye temperature $\Theta_{D,global} = (110 \pm 2)$ K, and $E_{a,global} = 0$ meV. Global fit quality $R$-square (COD) = 0.9997. (**a**) $\rho(T)$ data reported for thin film sample by Schenkel and Müller [27]. (**b**) $\rho(T)$ data reported for bulk sample sample by Schenkel and Müller [27]. (**c**) $\rho(T)$ data reported for thin film sample by Müller et al [22]. Other deduced parameters are shown in the figure. The 95% confidence bands are shown by pink shadow areas.

*3.8. Curium.* Figure 13 shows the $\rho(T)$ datasets for curium sample reported by Schenkel [26] and data fit with the NNH-parallel resistance model (Equation 10).

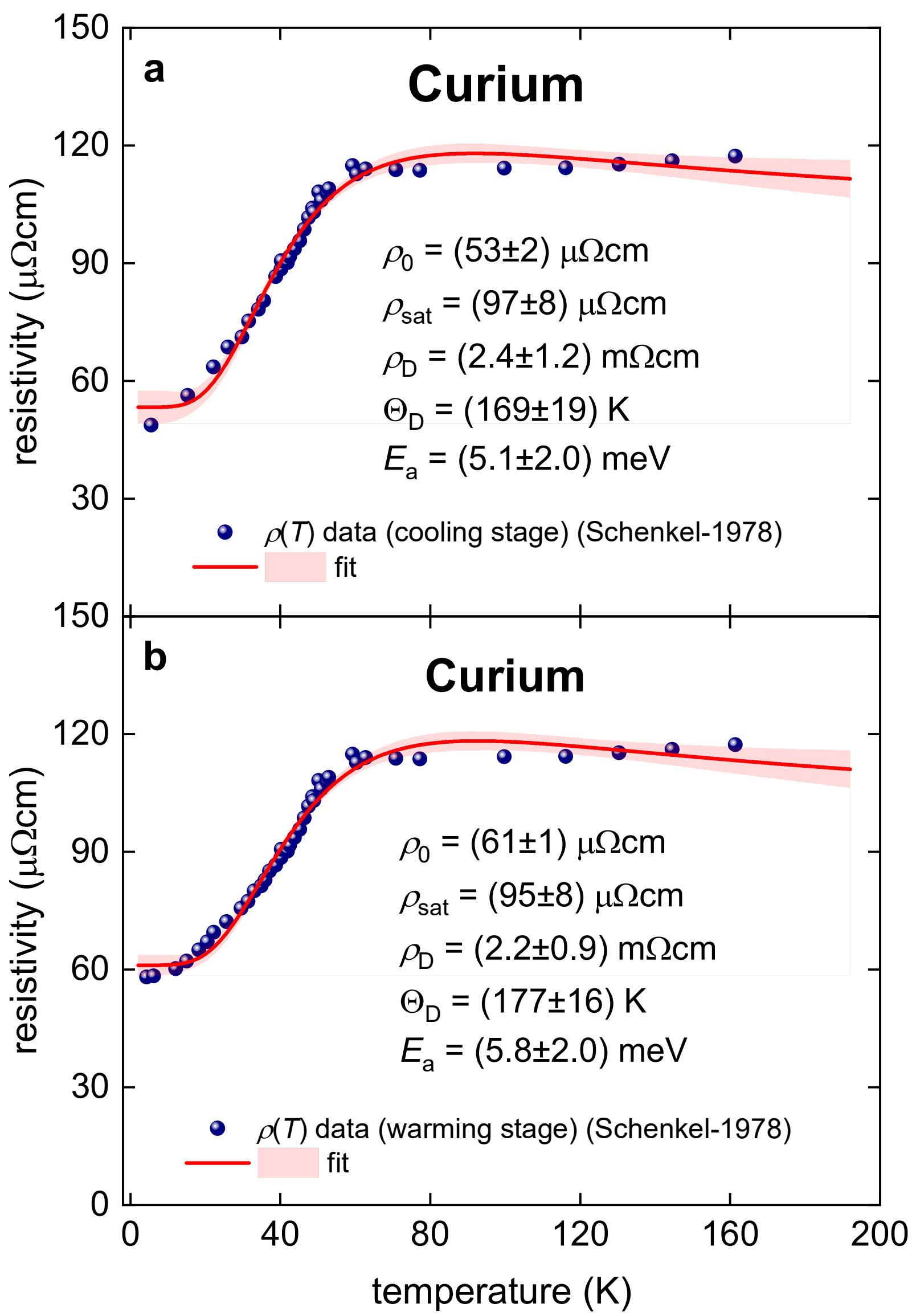

**Figure 13**. The fit using the nearest neighbor hopping parallel resistance model (Equation 10) to the $\rho(T)$ data for the curium sample reported by Schenkel [26]. (**a**) $\rho(T)$ data measured for the sample during the cooling stage. Deduced $E_a = (5.1 \pm 2.0)$ meV is equivalent to the temperature $\frac{E_a}{k_B} = (59 \pm 23)$ K. Derived Debye temperature $\Theta_D = (169 \pm 19)$ K. Fit quality $R$-square (COD) = 0.9836. (**b**) $\rho(T)$ data measured for the sample during the warming stage. Deduced $E_a = (5.8 \pm 2.0)$ meV is equivalent to the temperature $\frac{E_a}{k_B} = (67 \pm 24)$ K. Deduced Debye temperature $\Theta_D = (177 \pm 16)$ K. Fit quality $R$-square (COD) = 0.9842. Other derived parameters are shown in the figure. The 95% confidence bands are shown by pink shadow areas.

It is important to note that Figure 13 shows that the following parameters remained unchanged while the curium sample was cooled down (Figure 13,a) and warmed up (Figure 13,b):

1. Debye temperature $\Theta_D \cong 175$ K;
2. Arrhenius activation energy $E_a \cong 5.5$ meV;
3. Saturated resistance $\rho_{sat} \cong 96$ μΩcm;
4. Resistivity constant for the electron-phonon contribution $\rho_D \cong 2.3$ mΩcm.

These findings showed that the NNH model (Equation 10) is stable and, as expected from physics, the difference can occur only for the residual resistivity parameter $\rho_0$ during cooling-warming cycling. Based on that fundings, the global data fit (Figure 14) was performed with the mentioned above parameters settled as global free-fitting parameters, and the only residual resistivity $\rho_0$ parameter was a free-fitting parameter for both samples in this fit. One can see the free-fitting global parameter in Figure 14.

Deduced Debye temperature $\Theta_{D,global} = (180 \pm 11)$ K differs from the value of $\Theta_D = 121$ K reported in the only available paper [101], and thus more experimental datasets/studies are required.

*3.9. δ-phase of Pu alloys: Pu-Ce and Pu-Ce-Ga.* For completeness, in Figure 15 I show the applicability of the model to the $\rho(T)$ data measured in alloys exhibiting the δ-phase of Pu [43]: Pu-6.1 at.% Ce (Figure 15,a), Pu-8.1 at.% Ce (Figure 15,b), and Pu-4.0 at.% Ce-0.9 at.% Ga (Figure 15,c).

It can be seen (Figure 15) that the extracted Debye temperature for these alloys is constant, $\Theta_D \cong 175$ K, which is the expected result considering that the alloys have low levels of doping elements. Also, the fits reveal that the Arrhenius energy is nearly constant too $E_a \cong 17$ meV.

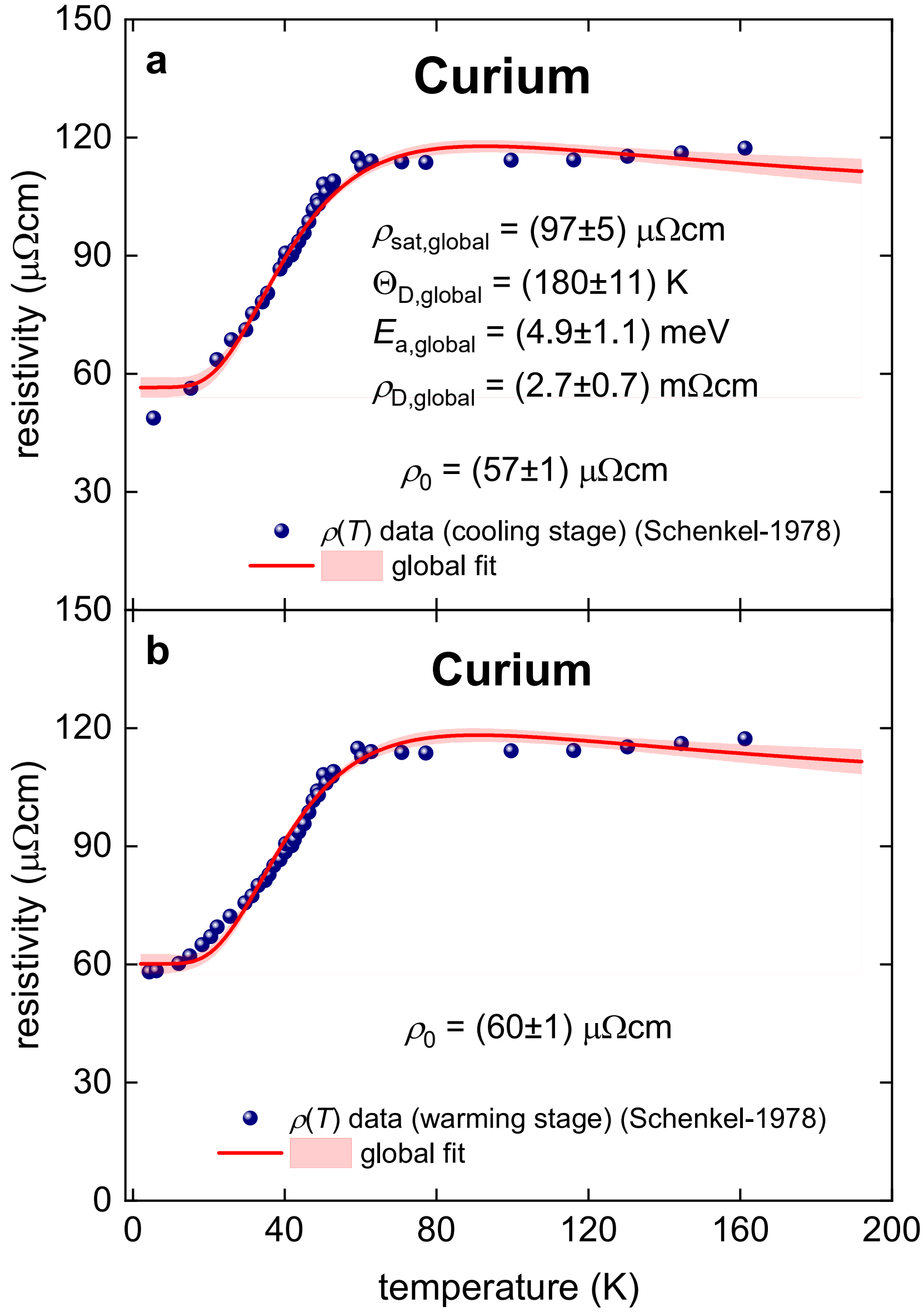

**Figure 14**. The global fit using the nearest neighbor hopping parallel resistance model (Equation 10) to the ρ(T) data for the curium sample reported by Schenkel [26]. Deduced $E_{\mathrm{a,global}} = (4.9 \pm 1.1)$ meV is equivalent to the temperature $\frac{E_a}{k_{\mathrm{B}}} = (57 \pm 14)$ K. Derived Debye temperature $\Theta_{\mathrm{D,global}} = (180 \pm 11)$ K. Other derived parameters are shown in the figure. Fit quality $R$-square (COD) = 0.9817. (**a**) $\rho(T)$ data measured for the sample during the cooling stage. (**b**) $\rho(T)$ data measured for the sample during the warming stage. The 95% confidence bands are shown by pink shadow areas.

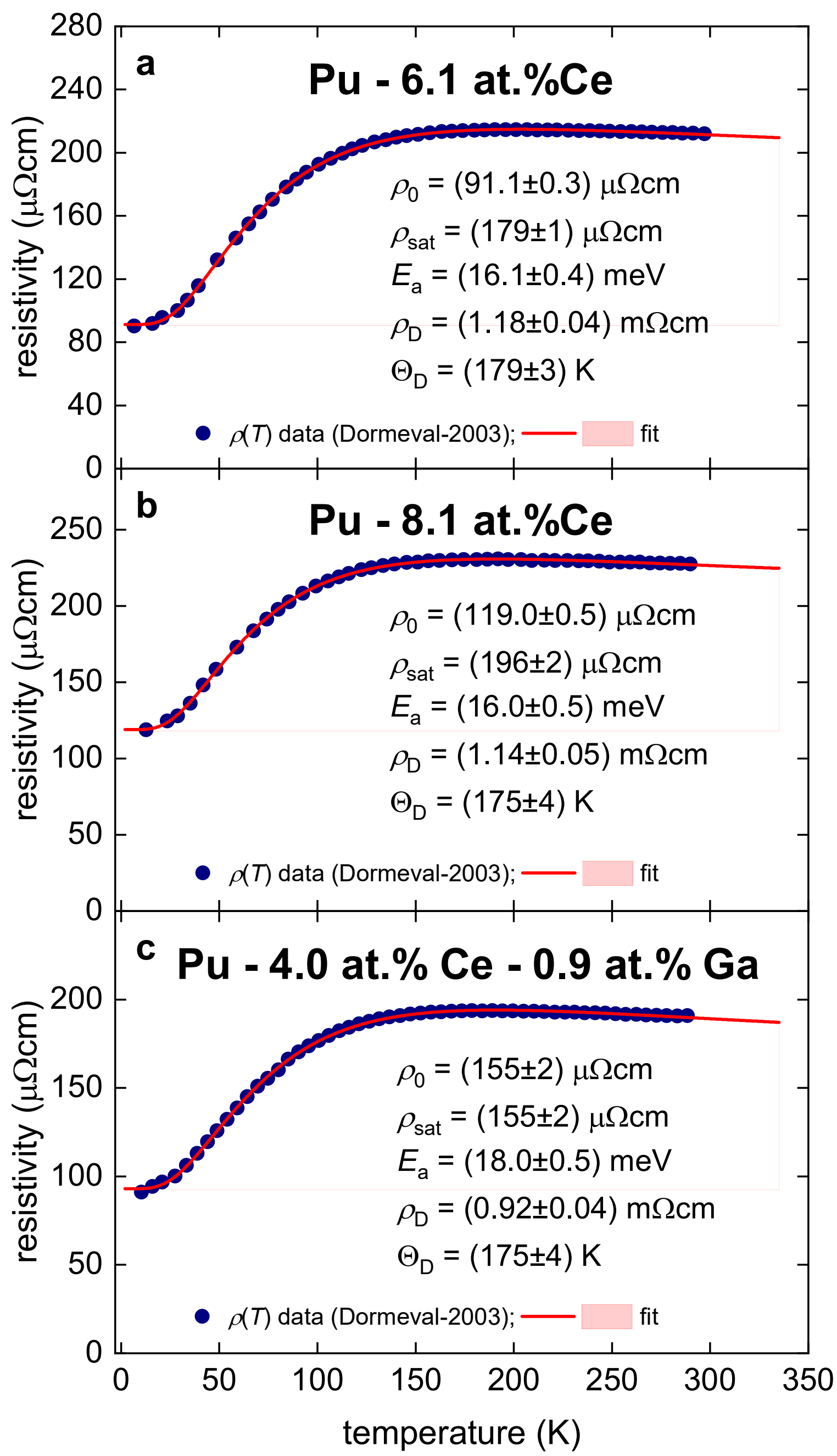

**Figure 15**. The fit using the nearest neighbor hopping parallel resistance model (Equation 10) to the ρ($T$) data for alloys having δ-phase of plutonium. Data reported by Dormeval et al [43]. (**a**) $\rho(T)$ data for Pu-6.1 at.% Ce alloy. Deduced $E_a = (16.1 \pm 0.4)$ meV, which is equivalent to the temperature $\frac{E_a}{k_B} = (186 \pm 4)$ K. Fit quality $R$-square (COD) = 0.9998. (**b**) $\rho(T)$ data measured for data for Pu-8.1 at.% Ce alloy. Deduced $E_a = (16.0 \pm 0.5)$ meV is equivalent to the temperature $\frac{E_a}{k_B} = (186 \pm 5)$ K. Fit quality $R$-square (COD) = 0.9842. (**c**) $\rho(T)$ data measured for data for Pu-4.0 at.% Ce-0.9 at.% Ga alloy. Deduced $E_a = (18.0 \pm 0.5)$ meV is equivalent to the temperature $\frac{E_a}{k_B} = (209 \pm 6)$ K. Fit quality $R$-square (COD) = 0.9996. Other derived parameters are shown in the figure. The 95% confidence bands are shown by pink shadow areas.

*4. Conclusions and Perspectives.* To the best of my knowledge, the theoretical idea that electrical resistivity in actinides is governed by a hopping mechanism was first expressed by Long [112] in 1977 (and from my understanding, the physics is described in details by Long [112]). And although hopping physics has been considered, along with other theoretical possibilities, for a long time since then [112,113], as the main mechanism for charge transfer in actinides, the model for the $\rho(T)$ data based on this mechanism has not been created.

One of the main problems associated with this originates (as I understand it) with the ongoing (decades-long) attempts to implement Matthiessen's rule (Equation 1) into the fitting equation for $\rho(T)$ in actinides and perhaps for other single-phase electrical conductors. These attempts were based on a fundamentally flawed concept, since in single-phase conductors electrical transport always occurs in parallel and, therefore, the primary equation is described by Equation 8.

This is fundamentally different from heat capacity models, where the total heat capacity is, in fact, the sum of the contributions from each channel [24,36,49,114]:

$$C_{\mathrm{p}}(T) = \sum_{i=1}^{N} C_{\mathrm{p},i}(T), \qquad (17)$$

where $C_{\mathrm{p},i}(T)$ represents a contribution from the *i*-th heat capacitance mechanism. But for the electrical resistivity, $\rho(T)$, in single-phase conductors, the main equation is based on the reciprocal values (Equation 8).

It should be also mentioned that the main concept described here, which is the hopping parallel resistance model (and that is Equations 10, 14 and 15), is more likely to have a wider application than has been explored to date.

Figure 16 summarized findings of this study for pure actinide elements.

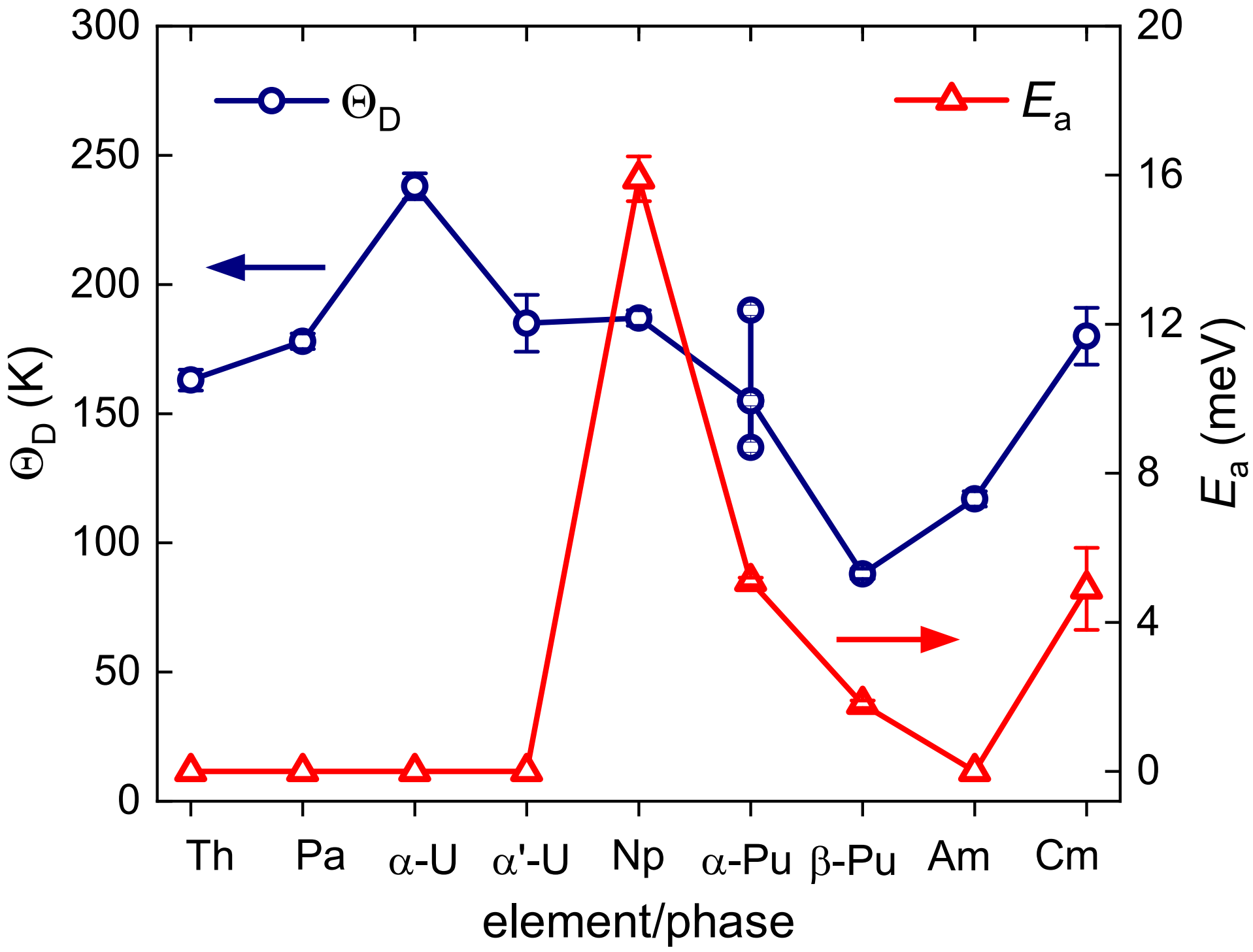

**Figure 16**. Parameters deduced by the nearest neighbor hopping parallel resistance model (Equations 10 and 16) for pure elemental actinides.

**Data availability statement**

The data that support the findings of this study are available from the corresponding author upon reasonable request.

**Declaration of interests**

The author declares that he has no known competing financial interests or personal relationships that could have appeared to influence the work reported in this paper.

**Acknowledgement**

The work was carried out within the framework of the state assignment of the Ministry of Science and Higher Education of the Russian Federation for the IMP UB RAS. The author gratefully acknowledges the research funding from the Ministry of Science and Higher Education of the Russian Federation under Ural Federal University Program of Development within the Priority-2030 Program.